\documentclass[11pt]{article}
\usepackage{moriond,epsfig}

\def\be{\begin{equation}}
\def\ee{\end{equation}}
\def\bea{\begin{eqnarray}}
\def\eea{\end{eqnarray}}

\def\mt1{\ensuremath{\tilde{m}_1}}
\def\MWR{\ensuremath{M_{W_R}}}

\begin{document}
\vspace*{4cm}
\title{Gauge dilution in leptogenesis}

\author{ N. Cosme }

\address{Service de Physique Th\'eorique, CP225\\
Universit\'e Libre de Bruxelles,\\
Bld du Triomphe, 1050 Brussels, Belgium.
}

\maketitle

\abstracts{We discuss leptogenesis in its natural context, i.e. gauge
unification, by including in the analysis the effects
of the gauge 
sector associated to the Majorana neutrinos. It results in stricter
bounds on the minimal Majorana mass, depending on the gauge bosons
mass, but also prevents to deduce any lower limit on the Yukawa
couplings since thermal and re-heating scenarios are now indistinguishable.}

\section{Introduction}

The baryogenesis through leptogenesis scenario is a very appealing
one in order to account for the observed baryon asymmetry in the
universe. 

In this scenario, heavy Majorana neutrinos linked to light left-handed
neutrinos via Yukawa couplings decay namely through channels violating
both lepton number and CP, once the temperature of the universe drops
below 
the Majorana scale. These Majorana neutrinos decay out of equilibrium
mainly according to the weakness of their Yukawa couplings. 

As a consequence, all Sakharov conditions may be fulfilled to create
a non zero lepton asymmetry.

Furthermore, around the electroweak phase transition, non-perturbative
$B-L$ conserving interactions, as exemplified by sphalerons, convert
this initial lepton number partially to  baryons giving thus rise to
the observed baryon asymmetry.

We will discuss this issue further in relation with
the neutrino mass parameters and with special emphasis on the effect
of gauge couplings to  Majorana neutrinos from the unification.
\cite{jmf&fsling}$^,$ \cite{Cosme:2004xs}
These gauge couplings are for us an integral part of the extention to right-handed
neutrinos but are usually (and unjustifiably) neglected.

\medskip

A simple model of leptogenesis is the standard model together with a
generation of right-handed neutrinos having a Majorana mass term. We list here the relevant lepton Yukawa couplings: 
\be \bar{l}_L \phi \lambda_l \; e_R + \bar{l}_L \tilde{\phi}
\lambda_\nu \; N +\frac{1}{2} \bar{N^c} M N +h.c.,\ee
where $l_L = \left( \nu_L, \; e_L \right)^t$, $\phi$
is the standard model scalar doublet and
$N$ the Majorana neutrinos with mass matrix $M$.

In addition to the above interactions, the unification is obviously
accompanied by gauge bosons coupled to Majorana
neutrinos. Irrespective of the details of the symmetry breaking, we
can expect that the induced Majorana mass will be related to the
masses of some gauge bosons by the breaking mechanism.

As a result, these degrees of freedom would be relevant for the
leptogenesis process and therefore a full description should include
the part of the gauge sector linked to right-handed neutrinos.

At the very least, one should consider right-handed charged and
neutral currents coupled to particles as $W_R^\pm$ and a $Z^\prime$
from an $SU(2)_R$ gauge symmetry.

While an explicit passage through the stage $SU(2)_R\times SU(2)_L
\times U(1)_{B-L}$ in the actual breaking of the unification structure
is not mandatory, the inclusion of the above-mentioned particles is
logical and characteristic of the actual Majorana neutrino gauge
couplings.

The inclusion of these gauge couplings in the description of
leptogenesis has consequences at different levels:
\begin{enumerate}
\item it dilutes the CP asymmetry by new CP conserving decay channels;
\item it reduces the Majorana decoupling from the thermal bath
through additional diffusions mediated by gauge bosons;
\item it favors the Majorana production if the Majorana population have
to be establish, e.g. in a re-heating scenario.
\end{enumerate}

\section{CP asymmetry}

The CP asymmetry in the decays of heavy Majorana neutrinos arises in
the channel ($N_i \to l_L \; \phi +  N_i \to \bar{l}_L \;
\phi^\dagger$) at one loop order from the interference of the tree level
amplitude with the vertex and self energy corrections:
\be \epsilon_i^\phi= \frac{\Gamma(N_i \to l \; \phi)-\Gamma(N_i
\to \bar{l} \; \phi^\dagger)}{\Gamma(N_i \to l \; \phi)+\Gamma(N_i \to
\bar{l} \; \phi^\dagger)}. \label{e1phi}\ee

If we assume hierarchical Majorana masses : $M_1<<M_2<<M_3$, the only 
$N_1$ decay is relevant regarding the final lepton asymmetry.
It is then possible to derive model independent upper bounds
on the CP asymmetry which will then be used as a necessary condition
for successful leptogenesis.
For instance, Davidson and Ibarra deduce the following upper bound
:\cite{DI} 
\be | \epsilon_1^\phi | \leq \epsilon_{DI}^\phi = \frac{3}{16 \pi}
\frac{M_1}{v^2} \left(m_3-m_1\right).\ee

This bound has however been further improved in \cite{Hambye} that we
will use in the following.

Including gauge interactions for the Majorana allows additional decay
channels which are CP conserving.

Two cases can be distinguished. First, when $M_1>\MWR$, that is the
$W_R$ gauge boson is lighter than all Majorana neutrinos, the
additional decay channels mostly occur in a two body decay into an
on-shell $W_R$. This possibility
implies a quite large partial decay width which dilutes excessively
the CP asymmetry leading to an unsuccessful leptogenesis scenario, and
is therefore excluded.\cite{jmf&fsling} 
In the second case, for $M_1<\MWR$, the
dominant additional channels are three body decays.  In this second
case, we keep the possibility of 
a successful leptogenesis. We will not consider the $M_1 \simeq \MWR$
degenerate case here.

\section{Diffusion Reactions}

Beside the decay processes leading to the generation of the lepton
asymmetry, we also have to consider the reactions constituted by
all the possible diffusions of Majorana neutrinos and leptons.

These diffusions keep the Majorana neutrinos coupled to the other
particles in the thermal bath and prevent for the departure from
equilibrium. 
In brief,  successful leptogenesis relies on the
strength of the different reactions. More accurately, the
dimensionless quantity involved in the evolution equations is the
ratio $\gamma/s$ to $H$, i.e. the ratio of the interaction rate per
unit of time to the universe expansion rate:\footnote{where
$\hat{\sigma}$ is the reduced cross section defined as
$\hat{\sigma}(aI \to J)(s) =(8/s) [(p_a . p_I)^2 -m^2_a m^2_I]
\sigma(s)$, $z=M_1/T$, $x=s/M_1^2$.
} 
\be
H(z) \propto \frac{M_1^2}{z^2\, M_{pl}};\quad s(z)\propto
\frac{M_1^3}{z^3},
\quad \gamma(z)\propto \frac{M_1^4}{z} \int dx \hat{\sigma}(x)
\sqrt{x} K_2(x\sqrt{z}).
\ee

The couplings in the scalar sector give rise to $\Delta L=1$
(e.g. $N_1 l_L \to \bar{t}_R Q_L$) and
$\Delta L=2$ (e.g. $ l_L l_L \to \phi \phi$) diffusion reactions. The
former are involved in the thermal coupling of Majorana neutrinos
while the latter induce a wash out of the lepton asymmetry as they are
in equilibrium. The dependence in the neutrino mass parameters is as follows:
\be
\frac{1}{H(M_1)s} \gamma_{\Delta L=1} \propto \mt1\;
\frac{M_{Pl} m_t^2}{v^4}, \qquad
\frac{1}{H(M_1)s}\gamma_{\Delta L=2} \propto \mt1^2 M_1
\;\frac{M_{Pl}}{v^4}, 
\label{gN}
\ee
where we restrict to the only $N_1$ Majorana in the propagator for the
second rate. We
introduced here the important parameters $\tilde{m}_1 = \frac{v^2 
\left( \lambda_\nu \lambda_\nu^\dagger\right)_{11}}{M_1}$.
Since it mimics a see-saw type formula, \mt1 provides
an effective light neutrino mass relevant in the estimate of the decay
width and of the other Majorana neutrino interactions.

Once we include the Majorana gauge interactions, we observe an
additional behavior along the Majorana mass parameter. Indeed, the
strength of $W_R$ and $Z^\prime$ mediated interactions (e.g. $N_1 e_R
\to \bar{u}_R d_R$, $N_1 N_1 \to e^-_R e^+_R$) fall when $M_1$
increases:\cite{Moha} 
\be \frac{1}{H(M_1) s}
\left( \gamma_{W_R} ,\gamma_{Z^\prime} \right) \propto \frac{1}{M_1}\,
M_{Pl}. \label{gWR}\ee


\begin{figure}[t!]
\begin{center}
\hspace{-.09 \linewidth}\includegraphics[width=.5\linewidth]{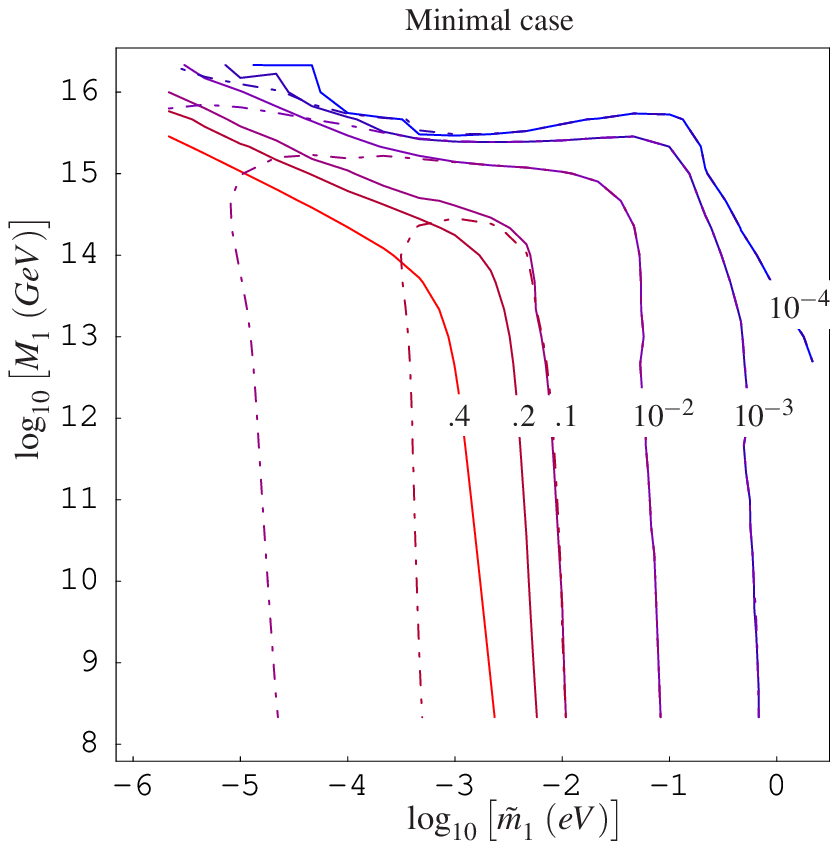}
\hspace{-.09 \linewidth}\includegraphics[width=.5
\linewidth]{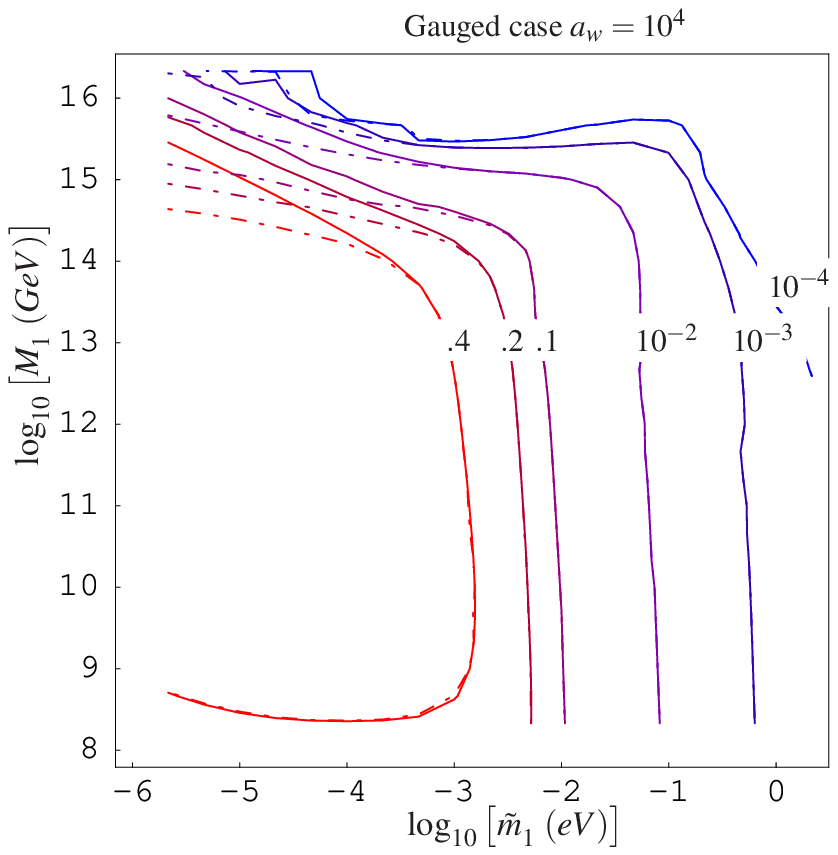}

\hspace{-.09 \linewidth}\includegraphics[width=.5 \linewidth]{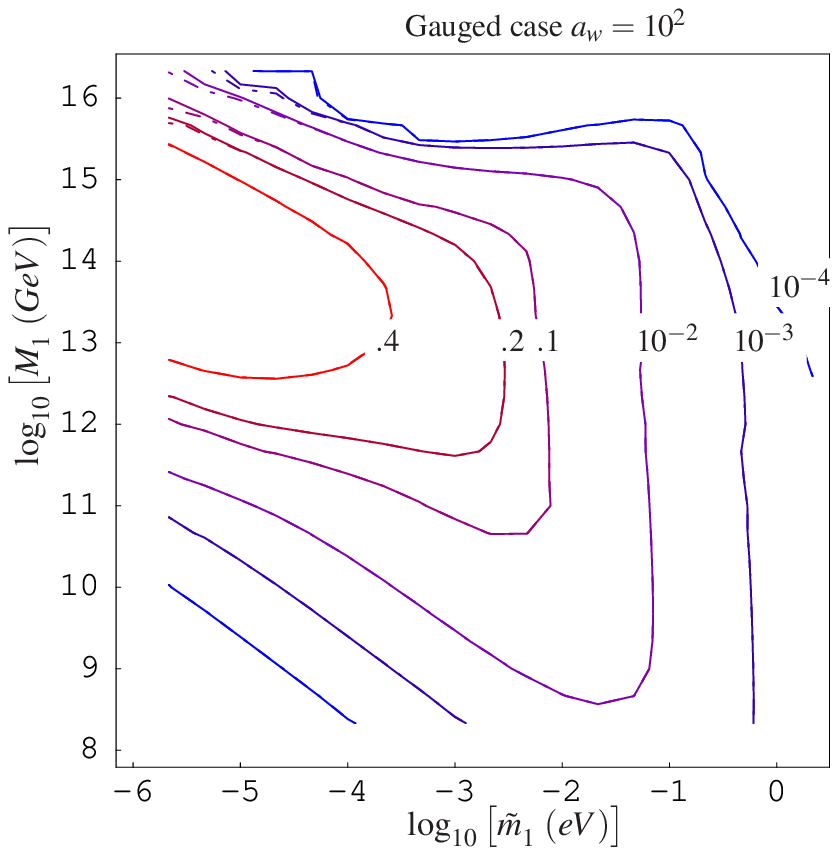}
\hspace{-.09 \linewidth}\includegraphics[width=.5 \linewidth]{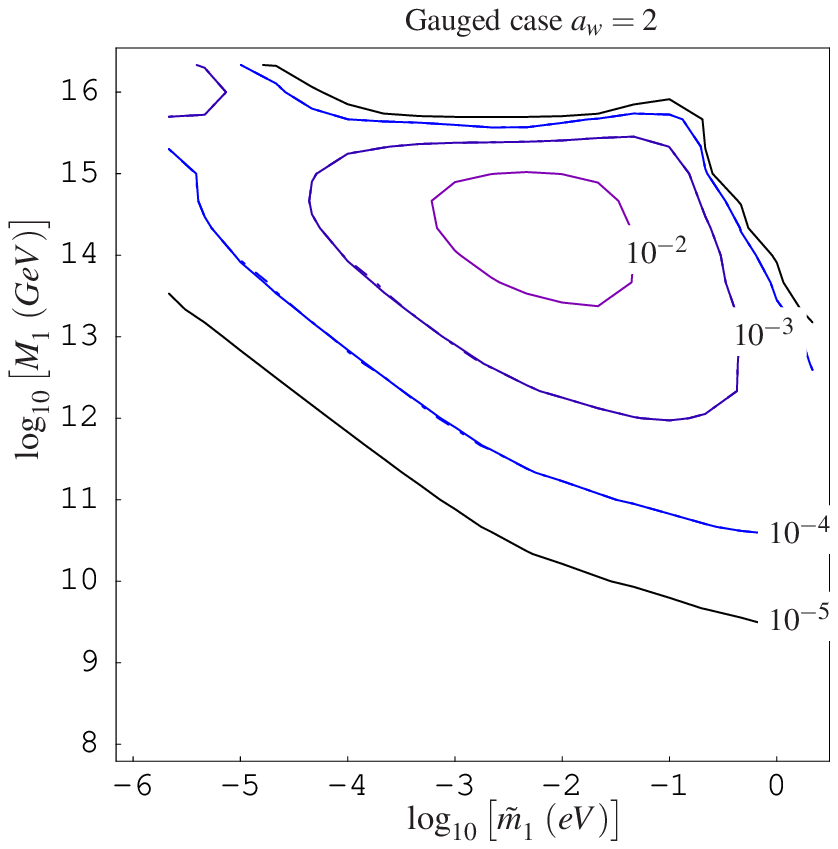}
\caption{Efficiency of leptogenesis for the minimal case $a_w \to \infty$ and
the extended case to a minimal right-handed gauge sector for $
a_w=\MWR^2/M_1^2 = 10^4, \; 10^2, \;2$.
The continuous line is the efficiency for a thermal initial Majorana abundance (thermal scenario) while the
dashed-doted line is for a zero initial Majorana abundance (re-heating scenario).}
\label{eta}
\end{center}
\end{figure}

The complete dynamics of the lepton number creation process may be
conveniently synthesized in a dimensionless quantity called the
efficiency which corresponds roughly to the percentage of lepton number
actually created in the process with respect to the maximum lepton
number possible.\cite{Barbieri}$^,$ \cite{Buch}$^,$ \cite{Giudice} We
present in Figure \ref{eta} the efficiency of  
leptogenesis in the $(\mt1,M_1)$ plane in the case where the gauge
interactions for Majorana neutrinos are neglected and for given $W_R$
to Majorana mass squared ratios $a_w=(M_{W_R}/M_1)^2$.

Different behaviors are revealed.
In the minimal case (corresponding to the artificial case $a_w \to
\infty$), two main effects are seen : First, for low $\mt1$, the
Majorana neutrino decays strongly out of equilibrium but in the
re-heating scenario (zero initial Majorana abundance) the Yukawa
couplings limit the Majorana production.
For larger $\mt1$,  the Majorana decays almost in
equilibrium resulting in less lepton number but the Majorana production
is then no more a problem. Secondly, for large $M_1$, the produced lepton
number is washed out by $\Delta L=2$ diffusions mediated by Majorana
neutrinos.

As we turn on the gauge interactions, this description is modified by
two points:

The creation of heavy Majorana neutrinos is enhanced by the gauge
interactions so that there is far less dependency according to the
initial condition, even for a sizable $a_w$ ratio.

The second consequence is a drastic reduction of leptogenesis
efficiency for low Majorana mass which increases dramatically for low
$a_w$ ratio. This comes from both delayed or impeded decoupling of
Majorana neutrinos through additional diffusions and decay channels.

Both effects impact the conclusions drawn for the
neutrino mass parameters. 

\section{Baryon asymmetry and neutrino mass}

Using the ability of leptogenesis to account for
the observed baryon number, we can provide constraints on the neutrino
mass parameters. These bounds are indeed useful for the understanding
of neutrino mass patterns and hence physics at energy well above the
standard model.

\begin{figure}[t!]\begin{center}
\hspace{-5cm}
\includegraphics[width=.8\linewidth]{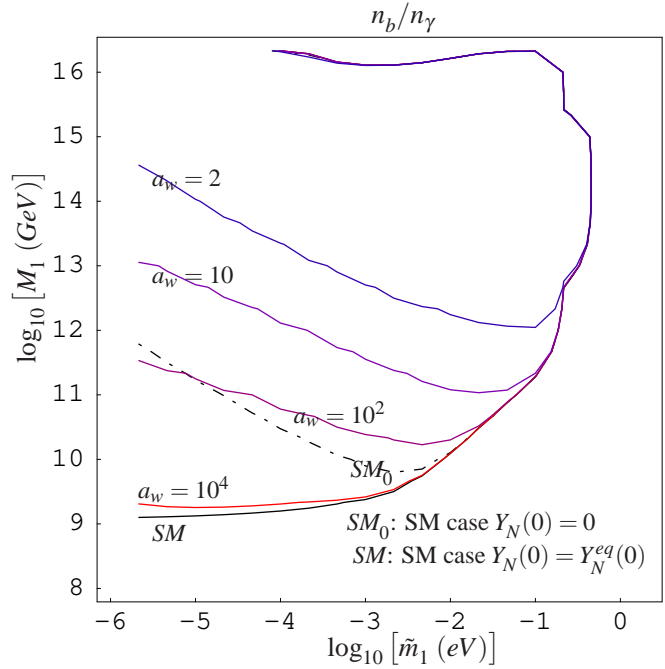}
\caption{Limits on the baryon to photon ratio in an extended model 
to a minimal right-handed gauge sector with
$a_w=\MWR^2/M_1^2 =10^4, \;\; 10^2, \;\;10, \;\; 2$,  from bottom to top (red to blue).
The minimal SM case is also shown in black:
the continuous line is for a thermal initial Majorana abundance while the
dashed-doted line is for a zero initial Majorana abundance.
The matter asymmetry is higher than the WMAP value in the region bounded by the curves.}
\label{nbtot}
\end{center}\end{figure}

Therefore the baryon to photon ratio expected from leptogenesis is
expressed in terms of the sphaleron conversion rates\cite{Shapo} and the
leptogenesis efficiency ($\eta_{eff}$):
\be 
\frac{n_B}{n_\gamma}\simeq \frac{s}{n_\gamma} C_{sph} |Y_L|\simeq
\frac{196}{79} \epsilon^\phi_1 Y_{N_1}^{eq}(init.) \eta_{eff},
\ee
where we assume that the
bound on the CP asymmetry derived in \cite{Hambye} is saturated.
Figure \ref{nbtot}
shows the iso-$n_B/n_\gamma$ curves in the $(\mt1,M_1)$ plane obtained
for different $a_w$ ratios, assuming the central value of baryon
asymmetry from WMAP: \cite{Spergel:2003cb}
$\frac{n_B}{n_\gamma}=(6.1^{+0.3}_{-0.2})\times 10^{-10}$. The baryon
asymmetry is higher than this value in regions bounded by the curves.

As already mentioned, the re-heating scenario is almost indistinguishable with the gauge
inclusion from the thermal scenario.
As a consequence, the full description of leptogenesis prevents to
deduce any lower limit on the effective neutrino mass $\mt1$ as occurs
in an incomplete analysis.
For comparison, the minimal (non-gauged Majorana) model curve for an
initial thermal Majorana abundance coincides in practice with the
curve for $a_w=10^6$ (the latter irrespective of the thermal or
re-heating scenario).

Finally, successful leptogenesis requires a large enough Majorana
mass. The gauge inclusion moreover teaches us that it evolves with the
Majorana to $W_R$ mass ratio.

\section{Conclusion}

We considered the description of leptogenesis in its natural context,
i.e. gauge unification, by including in the analysis the gauge bosons
coupled to the Majorana neutrinos. 

The main results are:
\begin{enumerate}
\item the $W_R$ mass has to be larger than the Majorana mass $M_1$ in
order to get a successful leptogenesis scenario.
\item the minimal allowed Majorana mass increases depending on the
associated gauge boson masses.
\item the re-heating scenario is almost indistinguishable in our
framework from the case of a thermal scenario, resulting in a broader
window for Yukawa couplings in the case of zero initial Majorana
abundance (re-heating scenario).
\end{enumerate}

\section*{Acknowledgments}

This work is supported in part by IISN, la Communaut\'{e} Fran\c{c}aise de Belgique (ARC),
 and the belgian federal government (IUAP-V/27).

\end{document}